\documentclass[seceq,epsfig]{ptptex}

\input{psfig}

\usepackage{graphicx}
\usepackage{wrapft}

\newbox\grsign \setbox\grsign=\hbox{$>$} \newdimen\grdimen \grdimen=\ht\grsign
\newbox\simlessbox \newbox\simgreatbox \newbox\simpropbox
\setbox\simgreatbox=\hbox{\raise.5ex\hbox{$>$}\llap
     {\lower.5ex\hbox{$\sim$}}}\ht1=\grdimen\dp1=0pt
\setbox\simlessbox=\hbox{\raise.5ex\hbox{$<$}\llap
     {\lower.5ex\hbox{$\sim$}}}\ht2=\grdimen\dp2=0pt
\setbox\simpropbox=\hbox{\raise.5ex\hbox{$\propto$}\llap
     {\lower.5ex\hbox{$\sim$}}}\ht2=\grdimen\dp2=0pt
\def\simgt{\mathrel{\copy\simgreatbox}}
\def\simlt{\mathrel{\copy\simlessbox}}

\def\tem{t_{\rm em}}
\def\tobs{t_{\rm obs}}

\def\Rtrail{R_{\rm trail}}
\def\Rb{R_\beta}

\def\Gn{\Gamma_n}

\def\Gej{\Gamma_{\rm ej}}

\def\brel{\beta_{\rm rel}}
\def\Grel{\Gamma_{\rm rel}}
\def\Erel{E_{\rm rel}}

\def\lbar{\lambda\llap {--}}
                                                                                
\def\be{\begin{equation}}
\def\ee{\end{equation}}

\def\dM{\dot{M}}
\def\dMeq{\dot{M}_{\rm eq}}

\def\taub{\tau_\beta}

\def\tcoll{t_{\rm coll}}

\def\Tmax{T_{\rm max}}

\def\dn{\dot{n}}
\def\lbar{\lambda\llap {--}}

\def\EF{E_{\rm F}}
\def\Gdec{\Gamma_{\rm dec}}

\title{
Theoretical Aspects of Gamma-Ray Bursts }

\author{
Andrei M. {\sc Beloborodov}
}

\inst{
Physics Department, Columbia University, 538 W. 120th Street, 
New York, 10027, USA
\\
}

\recdate{
}

\abst{
Cosmological GRBs are discussed with an emphasis 
on their plausible connection with black holes.
GRBs can be triggered by collapse of stellar-mass objects
that leads to formation of a black hole 
and a transient debris disk with a huge accretion rate.
The disk is believed to produce a relativistic jet (``fireball'') that 
expands and emits to infinity the observed burst of gamma-rays.
This accretion-jet picture is similar to quasars and X-ray binaries,
however, there are important differences: the physical conditions 
and the cooling mechanism in the disk are very different. The observed 
radiation is emitted when the expanding fireball becomes transparent,
at distances much larger than the Schwarzschild radius. 
The burst is then observed as a powerful relativistic explosion and 
the transient accretion disk in its center serves as a brief source of 
energy that drives the explosion. 
The explosion picture depends on the fireball nuclear composition which is 
shaped close to the black hole. A large amount of free neutrons survive 
till the emission phase and link the physics of the central 
engine to observed radiation.
}

\begin{document}

\maketitle


\section{GRB Explosions and Black Holes}

Gamma-ray bursts (GRBs) occur every day in the sky and last typically 
seconds or minutes. The energy spectrum of the burst peaks at 0.1-1~MeV, 
and the truly unique feature of this phenomenon is its huge luminosity: 
the energy comparable to (or even exceeding) a supernova is emitted in just 
seconds. By contrast, a normal supernova light peaks on a week timescale:
the exploding star has to expand enough to become transparent, so that
its thermal energy can be radiated away, and this takes $\sim 10^6$~s. 
The short duration of GRBs implies that the mass of the emitting material
$M$ is much smaller than a stellar mass. 

Moreover, $Mc^2$ is far below the burst energy $E$, so we deal with a 
{\it relativistic} explosion: the emitting plasma should
develop a Lorentz factor $\Gamma\sim E/Mc^2\gg 1$.
Detailed calculations show that the GRB durations can be reconciled 
with their energies if $\Gamma>100$~\cite{LS}. For such a high $\Gamma$ 
the observed duration $\tobs$ is strongly reduced by the Doppler effect 
compared to the emission in the local frame: 
$\tobs=(1-v/c)\tem\approx (\tem/2\Gamma^2)$.

The burst takes place in an ambient medium, and hence the relativistic
ejecta must decelerate as they sweep up enough ambient material.
The energy dissipated in this deceleration gives rise to the observed
afterglow emission. As $\Gamma$ decreases with increasing radius $R$ 
the afterglow radiation is emitted on longer timescales 
$\tobs\sim(R/c\Gamma^2)$ and in softer bands. 

Thus, a phenomenological picture of the GRB emerged: a highly relativistic 
ejecta (fireball) emits the prompt $\gamma$-ray emission of short duration, 
which is followed by the afterglow from the deceleration stage.
%
The main emission mechanism is synchrotron which requires the 
presence of magnetic fields and nonthermal electrons in the explosion. 
This aspect of GRB theory is poorly understood; it involves complicated 
plasma physics and contains a number of unknown parameters. 
Therefore, the derivation of the explosion parameters from the data is 
ambiguous and the nature of the circumburst medium (interstellar medium? 
progenitor wind?) is still disputed. It is also unclear whether the ejecta 
are dominated by baryons (flux of kinetic energy) or magnetic field 
(Poynting flux). 

There is a growing observational evidence that GRBs are produced by 
massive-star progenitors, most likely of Wolf-Rayet type. In some cases, 
a supernova component has been detected in the afterglow, with the most 
impressive case of GRB~030329.~\cite{S} The suspected supernovae are of 
type Ic, which are also associated with black hole formation.

Observational constraints on the trigger mechanism are inferred 
from the temporal variability and the energy budget of GRBs.
The prompt $\gamma$-radiation is strongly variable on timescales 
$\delta t$ ranging from milliseconds to the overall duration of the burst. 
In other words, the emitted radiation front is spatially inhomogeneous on 
scales $10^{7}\simlt c\delta t\simlt 10^{11}$~cm. If the front was emitted 
by the ejecta, they should have a radial structure 
on scales $10^{7}\simlt\delta R\simlt 10^{11}$~cm. 
The high-$\Gamma$ outflow is causally disconnected on scales 
$\delta R>R/\Gamma^2=10^7R_{12}(\Gamma/300)^{-2}$~cm, so it was likely
created inhomogeneous: the central engine 
was unsteady on timescales as short as milliseconds. This 
indicates that the size of the central engine does not exceed $10^7$~cm.
Known objects of this size, which are capable to release energies above 
$10^{51}$~erg, are black holes or neutron stars.

Most of the proposed scenarios are based on the gravitational source 
of energy.~\cite{M}
It can be a neutron-star merger, a collapse of 
a rotating massive star (``collapsar''), or a delayed collapse of a 
spinning neutron star (``supranova''), all of which lead to 
the formation of a black hole and an accretion disk of debris.
GRBs can also be produced by 
just born neutron stars, 
however, we shall focus here on the accretion scenario
as it is directly related to the subject of this meeting: black holes. 

A relativistic explosion will be produced if the accretion disk creates 
a relativistic jet. This seems likely to happen given that jets are 
observed in other black-hole objects --- X-ray binaries and quasars.
The jet can be fed by the Blandford-Znajek process that extracts the 
spin energy of the black hole. It can also be a MHD outflow 
from the disk~\cite{PMAB,MYKS}.
The precise mechanism still remains to be established, and this 
is especially difficult in the context of GRBs where the central engine 
is hidden from direct observation by its optically thick ejecta.

A different aspect of GRBs, which is easier to understand, is the 
nuclear history of the burst. It turns out to play important role in the 
overall picture of the explosion: the GRB resembles a huge neutron bomb. 
Below we discuss the physical conditions in GRB accretion disks, then 
address the nuclear aspect of this phenomenon and its connection with the 
mechanism of observed emission.


\section{Hyper-Accretion} 

A disk with accretion rate $\dM$ can power a relativistic outflow 
(fireball) with luminosity
\begin{equation}
  L=\epsilon_f\dM c^2\approx 10^{51}\left(\frac{\epsilon_f}{0.01}\right)
        \left(\frac{\dM}{10^{32}{\rm g/s}}\right) {\rm ~erg~s}^{-1},
\end{equation}
where $\epsilon_f$ is the efficiency of $\dM c^2$ conversion into a fireball.
This picture is similar to accreting black holes in X-ray binaries, 
and the standard theory of disk accretion applies\cite{PWF,NPK,KM,B1}. 
However, $\dM$ is some $15$ orders of magnitude higher, which leads to 
important differences:
\medskip

1. The disk, its corona, and the jet are in perfect blackbody state because 
of their high densities and temperatures. The rates of photon emission and 
absorption are huge, the radiation maintains detailed equilibrium and has 
a Planckian spectrum everywhere near the black hole. Given the liberated
power $L\sim 10^{52}-10^{54}$~erg~s$^{-1}$ and the size of the engine
$\sim 10^6-10^7$~cm, one finds the blackbody temperature, 
\begin{equation}
kT\sim 1-10{\rm ~MeV}. 
\end{equation}

2. At such temperatures, an equilibrium population of $e^\pm$ pairs must 
be present. The rates of pair annihilation and creation by photon-photon 
interactions are huge, and the $e^\pm$ maintain a perfect Fermi-Dirac 
distribution with the occupation number
\begin{equation}
\label{eq:fpm}
  f_\pm(E)=\frac{1}{\exp[(E\pm\mu)/kT]+1},
\end{equation}
where $\mu\equiv\mu_-=-\mu_+$ is the electron chemical potential
(thermodynamic equilibrium of $e^\pm$ with radiation implies $\mu_++\mu_-=0$.)
\medskip

3. The disk density $\rho\simgt 10^{10}$~g~cm$^{-3}$ is some 15 orders of 
magnitude higher than in X-ray binaries. Matter with such densities 
has the Fermi energy level
\begin{equation}
 \EF=3m_ec^2\left(\lbar^3Y_e\frac{\rho}{m_p}\right)^{1/3}\sim 10{\rm ~MeV}, 
\end{equation}
where $Y_e=n_p/(n_n+n_p)$ and the charged fraction of baryons and 
$\lbar=\hbar/m_ec$.
$\EF$ is comparable to the mean thermal energy of the electrons $3kT$ and
hence the GRB disks are mildly degenerate. The chemical potential 
is then given by
\begin{equation}
   \frac{\mu}{3kT}=\left(\frac{\EF}{3kT}\right)^3, \qquad \mu\simlt 3kT.
\end{equation}
\medskip
Even mild degeneracy 
suppresses significantly the positron density, $f_+/f_-<1$ (eq.~\ref{eq:fpm}),
because the levels of typical thermal energies $\sim 3kT$ are significantly 
occupied by the ambient $e^-$, and the creation of new pairs is limited to 
the free levels. So, the positron density in the disk is modest despite its 
high temperature.
\medskip


4. Both $3kT$ and $\mu$ exceed the difference between the 
neutron and proton rest-masses, $m_nc^2-m_pc^2=1.3$~MeV. Therefore the 
electrons are energetic enough for the 
neutronization reaction --- the $e^-$ capture onto protons.
$e^+$ capture onto neutrons also takes place,
\begin{equation}
\label{eq:reactions}
  e^-+p\rightarrow n+\nu, \qquad e^++n\rightarrow p+\bar{\nu}.
\end{equation}
These reactions rapidly convert protons into neutrons and neutrons
back into protons, and establish an equilibrium $Y_e=n_p/(n_n+n_p)$.
Since the positron density is suppressed by degeneracy,
the reaction $e^-+p\rightarrow n+\nu$ is preferential and the equilibrium
is shifted to a higher neutron density ($Y_e<0.5$). Precisely the same 
mechanism drives neutronization in the core of a supernova collapse, 
leading to formation of neutron stars. 
\medskip

5. The material around the black hole has a huge optical depth for photons, 
and radiation diffusion is completely negligible on the accretion timescale. 
The only cooling mechanism of the disk is neutrino emission. Reactions 
(\ref{eq:reactions}) are the main channels of neutrino emission (although 
there are also other channels, e.g. $e^++e^-\rightarrow \nu+\bar{\nu}$).
Thus, the $e^\pm$ capture reactions regulate both the temperature and 
the nuclear composition of the accretion disk.

 
\medskip

An upper bound on the disk temperature is derived from
assumption that the disk does not loose the dissipated energy and instead
traps it and advects. The advective flow at a radius $r$ has energy 
density~\cite{B1} $w\approx (3/8)(r_g/r)\rho c^2$, where $r_g=2GM/c^2$ is the 
Schwarzschild radius of the black hole.
The energy density in such an advective hot flow is dominated by radiation 
and weakly degenerate $e^\pm$, so that
\begin{equation}
\label{eq:Tmax}
   \frac{11}{4}a\Tmax^4\approx \frac{3r_g}{8r}\rho c^2,
\end{equation}
where $11/4$ accounts for the contribution of relativistic weakly degenerate
$e^\pm$. Neutrinos make a noticeable contribution to the energy density if
they are thermalized (self-absorbed), and then $\Tmax$ will be slightly lower.
Equation~(\ref{eq:Tmax}) yields
\begin{equation}
   k\Tmax\approx 13\left(\frac{\rho}{10^{11}{\rm g~cm}^{-3}}\right)^{1/4}
          \left(\frac{r}{3r_g}\right)^{-1/4} {\rm ~MeV}.
\end{equation}
The actual disk temperature $T$ can be lower because of the neutrino cooling,
whose efficiency depends on $\dM$.
%

One can show that the rates of reactions~(\ref{eq:reactions}) are 
sufficiently high to establish an equilibrium $Y_e$, i.e. the GRB accretion 
disks reach $\beta$-equilibrium. This is easy to see for disks cooled 
efficiently by neutrino losses.
Indeed, the mean energy of emitted neutrinos,
$E_\nu\simlt 30$~MeV, is below the liberated accretion energy per nucleon,
$\sim 100-300$~MeV, so the efficient cooling implies that more than one
neutrino per nucleon is produced, and hence the equilibrium $Y_e$ is 
achieved.

\begin{figure}[t!]
\centerline{\includegraphics[width=8cm]{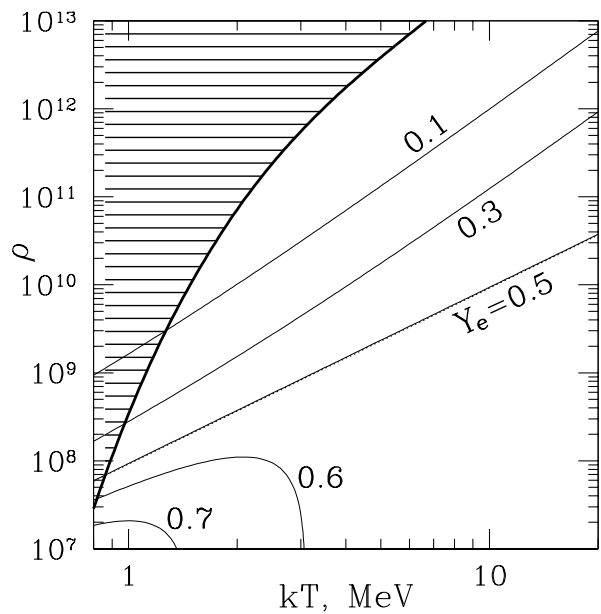}
}
\caption{Contours of equilibrium $Y_e=n_p/(n_n+n_p)$ on
the $T$-$\rho$ plane for $\nu$-transparent matter.~\cite{B1}
Composite nuclei dominate in the shaded region. 
} 
\end{figure}

In the opposite, advective, regime with $T\approx\Tmax$ the disk is
weakly degenerate, and the rates of $e^\pm$ capture in the zero order in 
$\mu/kT$ are
\begin{equation}
\label{eq:rates}
  \dn_{e^-p}\approx 1.5\times 10^{-2}n_p\theta^5, \qquad
  \dn_{e^+n}\approx 1.5\times 10^{-2}n_n\theta^5,
\end{equation}
where $\theta=kT/m_ec^2$.
The neutronization timescale is $t_n=n_p/\dn_{e^-p}\approx 70\theta^{-5}$~s
should be compared with the accretion timescale~\cite{B1} $t_a$, 
\begin{eqnarray}
\label{eq:ratio}
  \frac{t_n}{t_a}\approx
   2\times 10^{-2} \left(\frac{r}{3r_g}\right)^{13/8}
    \left(\frac{\alpha}{0.1}\right)^{9/4}\left(\frac{M}{M_\odot}\right)^{3/2}
    \left(\frac{\dM}{10^{32}}\right)^{-5/4}.
\end{eqnarray}
Disks with
\begin{equation}
\label{eq:dMeq}
  \dM>\dMeq\approx 10^{31}\left(\frac{r}{3r_g}\right)^{13/10}
  \left(\frac{\alpha}{0.1}\right)^{9/5}\left(\frac{M}{M_\odot}\right)^{6/5}
  {\rm ~g~s}^{-1}
\end{equation}
have $t_n<t_a$ and achieve the equilibrium $Y_e$; this range covers 
the plausible $\dM$ for GRBs.
Equation~(\ref{eq:dMeq}) as well as (\ref{eq:Tmax}) and (\ref{eq:ratio}) 
apply to both Schwarzschild and Kerr black holes; the relativistic 
corrections alter the expressions only slightly. The main effect of the 
black hole spin enters simply by decreasing the characteristic radius $r$ 
where most of accretion energy is liberated: $r$ decreases from 
$\sim 10 r_g$ (Schwarzschild) to $\sim r_g$ (Kerr).

The next question is very general: what is the 
equilibrium charged fraction $Y_e$ of matter with given $T$ and $\rho$?
The disk material should reach this $Y_e$ as it is in thermodynamic, 
nuclear, and $\beta$-equilibrium.
If the matter is transparent for neutrinos,
the equilibrium $Y_e$ is found by balancing 
the rates of reactions~(\ref{eq:reactions})~\cite{B1,PWH}. The 
result is shown in Fig.~1, and a simple analytical calculation shows that 
$Y_e<0.5$ if $\mu>Q/2$ where $Q=(m_p-m_n)c^2$.
If the matter is opaque to neutrinos,
$Y_e$ is found by balancing the chemical potentials 
$\mu+\mu_p=\mu_n+\mu_\nu$. Then $Y_e<0.5$ if $\mu>Q$. 

One concludes that accretion disks with $\mu>Q$ should have a neutron 
excess $n_n>n_p$. This condition translates to a condition for 
$\dM$,~\cite{B1}
\begin{equation}
\label{eq:dMn}
 \dM>\dM_n\approx 10^{31}
    \left(\frac{r}{3r_g}\right)^{3/2}\left(\frac{\alpha}{0.1}\right)
    \left(\frac{M}{M_\odot}\right)^2{\rm ~g~s}^{-1}.
\end{equation}
Plausible $\dM$ in GRB accretion flows are $10^{32}$~g/s and higher, and
they should be neutron rich.
This turns out important for the overall picture of the GRB explosion.



\section{Nuclear Composition of GRB Fireballs}

The baryonic component of the jet is picked up from the turbulent accretion
disk and remembers the disk $Y_e$, i.e. the jet has a lot of neutrons. 
The only threat to the escaping neutrons is the neutrino flux from the disk.
The timescale for $\nu$ absorption by $n$ (and $\bar{\nu}$ absorption by $p$) 
is shorter than the escape timescale if the neutrino luminosity 
$L_\nu>10^{53}$~erg/s. In that case, the neutrino flux 
controls $Y_e$ of the jet, which again is likely to give 
a neutron excess~\cite{Q,PFC}
(although not necessarily~\cite{B1}), and $Y_e$ freezes out quickly,
at $\sim 10$ Schwarzschild radii from the black hole.

The escaping jet is initially made of free nucleons $n$ and $p$, $e^\pm$ 
pairs, radiation, and magnetic field. The jet expands and cools adiabatically,
so that its internal energy is converted into bulk kinetic energy. 
When temperature drops to $\approx 10^2$~keV the free nucleons 
tend to recombine into $\alpha$-particles, i.e. nucleosynthesis is 
expected. The situation is very much similar to the big bang nucleosynthesis:
in both cases we deal with adiabatic expansion of radiation-dominated 
blackbody plasma. However, the outcome of nucleosynthesis turns out to be 
different~\cite{L,PGF,B1}.

Both the big bang and the GRB fireball can be described by 
three physical parameters: photon-to-baryon ratio $\phi=n_\gamma/n_b$, 
expansion timescale during the nucleosynthesis $t_*$ (measured in the
comoving frame), and $Y_e$, see the table below.
The table also shows the characteristic nucleosynthesis temperature
$T_*$ and the ratio of the recombination rate to the expansion rate at $T_*$. 
This ratio is $\sim 1$ for GRBs, which implies a marginally efficient 
nucleosynthesis, and the outcome depends on the precise values of parameters. 
An example is shown in Figure~2 with $\phi=10^5$, $r_0/c=10^{-4}$~s, $Y_e=0.5$.
%
The fireball is assumed to expand radially in this example.
A realistic GRB jet can be non-radial: it can be collimated 
by a parabolic funnel of the rotating and collapsing star or
by its own magnetic field~\cite{VPK}.
The nuclear reactions in a collimated jet have also been calculated~\cite{B1}.
The collimation generally helps nucleosynthesis because it increases
the expansion timescale, and then a significant fraction of nucleons 
recombine into $\alpha$-particles. 

\bigskip
\begin{center}
\begin{tabular}{cccccc}
\hline
&&&&&\\
     &  $\phi=n_\gamma/n_b$ & expansion timescale $t_*$ & $Y_e$ & $T_*$ & 
   $\frac{\rm recombination~rate}{\rm expansion~rate}$ \\ 
&&&&&\\
\hline
&&&&&\\
 Big Bang  & $10^{10}$ & $10^2$~s & $7/8$ & 80~keV & $\sim 10$\\
&&&&&\\
 GRB & $10^{5}$ & $10^{-4}$~s & $<1/2$ & 140~keV & $\sim 1$ \\
&&&&&\\
\hline
\end{tabular}
\end{center}
\bigskip

Even at conditions extremely favorite for nucleosynthesis, when the 
recombination 
rate is much higher than the expansion rate, there are leftover 
neutrons in the fireball because of the neutron excess $Y_e<0.5$ (the 
formation of $\alpha$-particles consumes equal numbers of $n$ and $p$). 
The mass fraction of leftover 
neutrons is then $X_n=1-2Y_e$. For a similar reason, there are leftover 
protons in the big bang nucleosynthesis where $Y_e\approx 7/8>0.5$; the mass 
fraction of leftover protons $X_p=1-2(1-Y_e)=3/4$ defines the hydrogen fraction 
of the observed universe, and the remaining 1/4 is made of the recombined 
nucleons (predominantly $\alpha$-particles and a tiny fraction of other 
light elements). By contrast, the post-nucleosynthesis GRB fireball is most 
likely composed of neutrons and $\alpha$-particles. Another interesting 
difference from the big bang is the large amount of deuterium in the 
freezout (Fig.~2).

\begin{figure}[t!]
\centerline{\includegraphics[width=8cm]{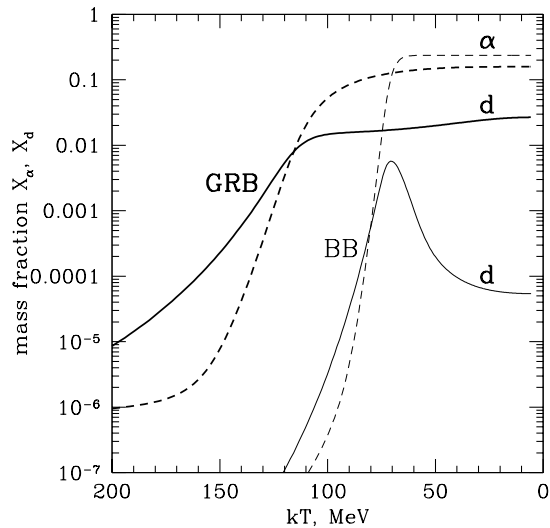}}
\caption{Evolution of deuterium ($d$) and helium ($\alpha$) abundances with
temperature in an expanding fireball. The (thicker) GRB curves are
calculated for a radial explosion with $\phi=n_\gamma/n_b=10^5$, $Y_e=0.5$ 
and $r_0=3\times 10^6$~cm. For comparison, the big bang (BB) nucleosynthesis
is also shown (with $\phi=3\times 10^9$). 
} 
\end{figure}


\section{Decoupling of Neutrons}

The fireball is accelerated by the radiation pressure and its Lorentz 
factor $\Gamma$ grows as long as $aT^4>n_bm_pc^2$, where $n_b$ is baryon 
density in the comoving frame. The neutron component practically does not 
participate in electro-magnetic interactions, however, the neutrons also 
accelerate because they are collisionally coupled (via strong interactions) 
to the ions.
The collisional timescale $\tcoll$ at small radii is very short compared 
to the fireball expansion timescale $R/c$, and
the neutron Lorentz factor $\Gamma_n$ only slightly lags behind the ion
Lorentz factor $\Gamma$: $(\Gamma-\Gamma_n)/\Gamma=\tcoll c/R <1$~\cite{DKK}.
During the fireball acceleration,
$n_b$ scales as $\Gamma^{-3}$
and the neutrons decouple ($\tcoll c/R >1$) before the acceleration ends
if $\Gamma$ reaches $\Gdec$
\begin{equation}
\label{eq:Gdec}
  \Gdec\approx\left(\frac{\sigma_0\dM_\Omega}{m_pcr_0}\right)^{1/3}
       \approx 300\left(\frac{\dM_\Omega}{10^{26}{\rm g~s}^{-1}}\right)^{1/3}
              \left(\frac{r_0}{3\times 10^6{\rm cm}}\right)^{-1/3}.
\end{equation}
Here $\sigma_0\approx 3\times 10^{-26}$~cm$^2$ describes the neutron-ion 
collisions, $r_0$ is the size of the central engine, and $\dM_\Omega$ is 
the outflow rate of baryonic rest mass per unit solid angle.
In radially expanding fireballs, $\Gamma(R)=R/r_0$;
this gives the decoupling radius $\Gdec r_0\sim 10^9$~cm. 
Magnetically collimated fireballs can be non-radial out to very large radii
\cite{VPK}, then $\Gdec$ can be one order of magnitude smaller.



Before the decoupling, the small relative velocity of the neutron 
fluid with respect to the 
ions, $\brel=\tcoll c/R<1$, leads to a significant heating of the fireball. 
The neutron component has the kinetic energy $\Erel=M_nc^2\brel^2/2$ in the 
ion fireball frame, which is dissipated on a short timescale $\tcoll$. 
On the other hand, $\brel$ is constantly pumped by the acceleration of the 
fireball frame.
The dissipation is similar to Ohmic heating in a conductor. Instead of 
electric current driven by an electric field, here the inertial 
``neutron current'' is caused by acceleration of the fireball frame.
Since the acceleration is driven by radiation, the dissipation in essence
converts a fraction of the radiation energy into the thermal energy of 
baryons.

The baryonic component is heated with 
rate $\Erel/\tcoll=(1/2) M_nc^2 (\tcoll c^2/ R^2)$.
%
If the heat remained stored in the baryons, the baryonic temperature 
$T_b(=T_{\rm ion}=T_n)$ would be $\approx m_pc^2 \brel$. 
However, the ions are thermally coupled to the electrons by Coulomb 
collisions, which are in turn coupled to radiation by Compton scattering.
Therefore, most of the dissipated heat flows back to radiation, which strongly 
dominates the heat capacity of the fireball ($n_\gamma/n_b\sim 10^5$). 
Only a fraction of the dissipated heat is kept by baryons to maintain 
$T_b$ sufficiently high above $T=T_e=T_\gamma$ so that the quasi-steady 
energy circulation is maintained in the accelerating fireball:
radiation $\rightarrow$ $\Erel$
$\rightarrow$ baryonic heat $\rightarrow$ radiation.

The baryonic temperature
$T_b$ is found by equating the heating rate to the rate of Coulomb exchange 
between the ions and electrons. For radial fireballs ($\Gamma=R/r_0$) 
this gives
\begin{equation}
   T_b\approx 10^{10} \left(\frac{\Gamma}{\Gdec}\right)^{9/2} 
   \left(\frac{\dM_\Omega}{10^{26}{\rm g~s}^{-1}}\right)^{-1/2}
              \left(\frac{r_0}{3\times 10^6{\rm cm}}\right)^{1/2}
               \left(\frac{T_0}{10^{10}{\rm K}}\right)^{3/2} {\rm ~K},
\end{equation}  
where $T_0=T(r_0)$ is the initial temperature of the fireball near the central
engine. This expression is valid at $\Gamma>(1/3)\Gdec$ where 
$T_b>T_0/\Gamma=T$. $T_b$ reaches values $\sim 10^{10}$~K, which is 
$10^2-10^3$ times higher than it would be in the 
absence of neutrons: a fireball without neutrons would be in the blackbody
state with $T_b=T=T_0/\Gamma$.


When $\tcoll$ approaches $R/c$ the neutrons decouple from the ions.
If the ion fireball is still accelerated by radiation at this moment,
the relative velocity of neutrons approaches the speed of light and their 
last collisions with the ions 
are very energetic, above the threshold for pion production. 
The produced pions decay into muons, which 
in turn decay into electrons and neutrinos. Thus, an observable 
flux of multi-GeV neutrinos is produced~\cite{DKK,BM}.

\section{$\beta$-Decay Inside the Fireball in the Prompt Emission Phase}

After the decoupling, the neutrons coast with a constant Lorentz factor
$\Gamma_n=const$. The ion fireball with energy $\eta m_pc^2$
per baryon accelerates to $\Gamma=\eta$ and, if $\eta>\Gdec$, the final
result of acceleration is $\Gamma>\Gn=\Gdec$. The fireball is then composed 
of two non-interacting components with different Lorentz factors.

The neutron component of total mass $M_n$ gradually decays with rate 
$\dM_n=M_nc/\Rb$ where 
\begin{equation}
 \Rb=c\taub \Gn=8 \times 10^{15}\left(\frac{\Gn}{300}\right) {\rm ~cm}
\end{equation}
is the mean decay radius and $\taub\approx 900$~s is the mean life-time 
of neutrons in their rest frame.
At radii $R<\Rb$ the amount of decayed neutrons $\Delta M_n=(R/\Rb)M_n$ 
is small, however, their impact on the fireball can be significant if 
$\Gamma>\Gn$.
Then the charged products of $\beta$-decay $p$ and $e^-$ have 
a significant Lorentz factor with respect to the ion fireball, 
$\Grel\sim\Gamma/\Gamma_n$, and they immediately share 
their momentum as the two-stream plasma instability damps the relative 
motion on a short timescale.\footnote{Another, and likely dominant, mechanism 
of momentum sharing is the gyration of the decay products in a transverse 
magnetic field frozen in the fireball (the transverse component is 
dominant in a radially expanding plasma as follows from magnetic flux 
conservation).}
The momentum exchange between the decayed neutrons and the ions 
is described as inellastic collision that 
heats up and decelerates the ion fireball. Calculations~\cite{RBR1} show that
at radii $10^{14}-10^{16}$~cm the ion temperature rises to 
$10^{11}-10^{12}$~K and the ion $\Gamma$ decreases.

Internal shocks are thought to occur in the fireball at about the same 
radii. The shocks develop because $\eta$ is likely inhomogeneous in the 
fireball and different parts of it coast with different $\Gamma$. The
shocks have been proposed as a mechanism of the prompt GRB emission.
The heating and deceleration of the high-$\Gamma$ parts of the fireball
by $\beta$-decay should reducing their Mach number. As a result, the 
amplitude of internal shocks is reduced or even suppressed completely, 
and this gives the most 
constraining upper bound on their efficiency.~\cite{RBR1}


\section{$\beta$-Decay in the Afterglow Phase}

 
The $\beta$-decay depletes exponentially the neutron component outside
the mean-decay radius $\Rb\approx 10^{16}(\Gn/300)$~cm, which is
comparable to the radius of the early afterglow emission.
It turns out that the neutrons have a huge impact on the external
blast wave at radii significantly larger than $\Rb$, even though their 
number is exponentially reduced.

At typical afterglow radii $10^{16}-10^{17}$~cm the fireball can be 
viewed as a very thin shell (thickness $\Delta\sim 10^{11}{\rm ~cm}\ll R$).
The survived neutron component coasts with a constant $\Gn$ while 
the ion component is decelerated by the external medium and eventually
lags behind the neutrons. When $\Gamma$ of
the ions drops below $\sim (R/\Delta)^{1/2}\sim$ a few hundred,
the fireball splits into two distinct shells: the leading neutrons and
the trailing ions.
 
The front of survived neutrons
overtakes the decelerating external shock wave at radius $R_*$ where the
shock-wave Lorentz factor $\Gamma$ decreases below $\Gn$. Using the
Blandford-McKee solution $\Gamma^2(R)=(17-4k)E/8\pi \rho_0 c^2R^3$
for adiabatic blast waves in a medium with density
$\rho_0\propto R^{-k}$ we find $R_*$ from equation
\begin{equation}
 R_*^3\rho_0(R_*)=\frac{(17-4k)E}{8\pi c^2\Gn^2}.
\end{equation}
A typical afterglow model with $\rho_0=const\sim 10^{-24}$~g~cm$^{-3}$,
$E\sim 10^{52}$~erg, and $\Gn\approx 300$, gives $R_*\sim 3\times 10^{16}$~cm.
At $R>R_*$ the $\beta$-decay takes place in the external medium {\it ahead} 
of the forward shock that produces the afterglow radiation.

The impact of this decay can be understood by comparing the energy of
neutrons, $E_n\approx X_nE\exp(-R/\Rb)$ ($X_n$ is the initial neutron 
fraction of the explosion) with the ambient mass
$mc^2=\frac{4\pi}{(3-k)}R^3\rho_0c^2=\frac{(17-4k)}{2(3-k)}(E/\Gamma^2)$
they interact with,
\begin{equation}
   \frac{E_n}{mc^2}=\frac{2(3-k)}{17-4k}X_n\Gamma^2
                     \exp\left(-\frac{R}{\Rb}\right).
\end{equation}
Immediately after $R_*$ this ratio can be as large as $\Gn^2\sim 10^5$, 
depending 
on $R_*/\Rb$. The decaying neutron front with $E_n>mc^2$ deposits huge 
momentum and energy into the ambient medium, leaving behind a relativistic
trail. The exact parameters of this trail are found from energy and momentum
conservation.~\cite{B2}.
                                                                                
The ratio $E_n/mc^2$ decreases to unity after
$\approx 10$ e-foldings of $\beta$-decay. Therefore the impact of neutrons
lasts until $\Rtrail\approx 10\Rb\approx 10^{17}$~cm, and
one expects an observational effect if $R_*<\Rtrail$. For
homogeneous medium ($k=0$) this requires a number density
$n_0>0.1 E_{52}(\Gamma_n/300)^{-5}$~cm$^{-3}$.
For wind-type models ($k=2$) $R_*<\Rtrail$ for all
plausible parameters of the wind if $\Gn\sim 10^2$ or higher.\footnote{ 
Besides, the forward shock 
in the dense wind of a Wolf-Rayet progenitor is likely to be slow from the 
very beginning (the reverse shock in the ejecta is relativistic, 
$\Gej\gg\Gamma$). Then $\Gamma<\Gn$ and the neutrons overtake the shock 
immediately, before the self-similar deceleration sets in.} 
                                                                                
The $\beta$-decay ahead of the shock transforms the cold static external 
medium into a hot, dense, relativistically moving, and possibly magnetized,
material. This transformation of the preshock medium should
affect the appearance of the afterglow radiation. 
Like the neutron-free shocks, it is difficult to 
calculate the emission from first principles because the 
electron acceleration and magnetic field evolution are poorly 
understood. The best one could do is to apply a phenomenological shock 
model with customary parameters $\epsilon_e$ and $\epsilon_B$:
$\epsilon_e$ is the fraction of shock energy passed to the nonthermal 
electrons and $\epsilon_B$ is the energy fraction in postshock magnetic field.  
This may enable an observational test for the $\beta$-decay in the 
afterglow phase.

\section{Conclusions}

The GRB phenomenon can be associated with formation of stellar-mass
black holes with small-scale disks of dense material that accretes quickly
and produces a relativistic jet.
The neutronization process takes place in the disk and, as a result,
the baryonic component of the jet is neutron rich.
The observed explosion then resembles a huge neutron bomb. 
The jet accelerates to Lorentz factors $10^2-10^3$, so the decay
time of neutrons is increased by the factor $10^2-10^3$ and an 
interesting fraction of neutrons survive out to large distances 
$\sim 10^{17}$~cm. Such distances cover the prompt phase of $\gamma$-ray 
emission and at least the early afterglow. The survived fraction of neutrons 
should overtake the 
external shock wave and deposit huge energy in the ambient medium, thus 
changing the very mechanism of the GRB blast wave.

Relativistic neutron outflows were also proposed to take place in 
AGN.~\cite{SBR} They originated from nonthermal protons in accretion disks 
around supermassive black holes and decayed far away in the ambient medium. 
In contrast to GRBs, the AGN neutron outflows are steady and uncollimated. 
The GRB neutrons have a better chance to be observed as they affect the 
observed development of the explosion.

Any neutron signature revealed in a GRB afterglow emission 
would confirm that the ejected baryonic material has gone 
through a hot high-density phase in the central engine. 
Neutrons thus provide a unique link between the
physics of the central engine and the observed afterglow.
Numerical simulations of neutron-fed blast waves may help to identify
such signatures. One possibility, for instance, is an exponentially
decaying emission component.
Another possible signature is a spectral transition or a bump in the 
afterglow light curve at $R\approx\Rtrail$~\cite{B2}.

Absence of neutron signatures would indicate that the GRB jets are 
dominated by magnetic fields. In such a low-$\dM_\Omega$ jet, neutrons 
would decouple early with a modest Lorentz factor (eq.~\ref{eq:Gdec}) and 
decay quickly. A two-component jet with less collimated and less energetic
neutrons is possible in the MHD acceleration scenario~\cite{VPK}.


\section*{Acknowledgements}
This research was supported by NASA grant NAG5-13382.


\begin{thebibliography}{99}
\bibitem{LS}
Y.~Lithwick and R. Sari, ApJ, {\bf 555} (2001), 540.

\bibitem{M}
P.~M\'esz\'aros, ARAA, {\bf 40} (2002), 137.

\bibitem{S}
K.Z.~Stanek et al., ApJ, {\bf 591} (2003), L17.

\bibitem{PMAB}
D.~Proga, A.I.~MacFadyen, P.J.~Armitage, and M.C.~Begelman, ApJ, {\bf 599}
(2003), L5 

\bibitem{MYKS}
Y.~Mizuno, S.~Yamada, S.~Koide, and K.~Shibata, ApJ, submitted 
(astro-ph/0404152)

\bibitem{PWF}
R.~Popham, S.E.~Woosley, and C.~Fryer, ApJ, {\bf 518} (1999), 356.

\bibitem{NPK}
R.~Narayan, T.~Piran, and P.~Kumar, ApJ, {\bf 557} (2001), 949.

\bibitem{KM}
K.~Kohri, and S.~Mineshige, ApJ, {\bf 577} (2002), 311.

\bibitem{B1}
A.M.~Beloborodov, ApJ, {\bf 588} (2003), 931.

\bibitem{PWH}
J.~Pruet, S.E.~Woosley, \& R.D.~Hoffman, ApJ, {\bf 586} (2003), 1254.

\bibitem{Q}
Y.-Z.~Qian, G.M.~Fuller, G.J.~Mathews, R.W.~Mayle, J.R.~Wilson, and
S.E.~Woosley, Phys. Rev. Lett., {\bf 71} (1993), 1965.

\bibitem{PFC}
J.~Pruet, G.M.~Fuller, and C.Y.~Cardall ApJ, {\bf 561} (2001), 957.

\bibitem{L}
M.~Lemoine, A\&A, {\bf 390} (2002), L31.

\bibitem{PGF}
J.~Pruet, S.~Guiles, and G.M.~Fuller, ApJ, {\bf 580} (2002), 368.

\bibitem{VPK}
N.~Vlahakis, F.~Peng, and A.~K\"onigl, ApJ, {\bf 594} (2003), L23. 

\bibitem{DKK}
E.V.~Derishev, V.V.~Kocharovsky, and Vl.V.~Kocharovsky, ApJ, {\bf 521} (1999),
640.

\bibitem{BM}
J.N.~Bahcall and P.~M\'esz\'aros, Phys. Rev. Lett., {\bf 85} (2000), 1362.

\bibitem{RBR1}
E.M.~Rossi, A.M.~Beloborodov, and M.J.~Rees (2004) in preparation


\bibitem{BlM}
R.D.~Blandford and C.F.~McKee, Phys. Fluids, {\bf 19} (1976), 1130.

\bibitem{B2}
A.M.~Beloborodov, ApJ, {\bf 585} (2003), L19.

\bibitem{SBR}
M.~Sikora, M.C.~Begelman, and B.~Rudak, ApJ, {\bf 341} (1989), L33

\end{thebibliography}
\end{document}